\documentclass[12pt,preprint]{aastex}

\def\la{\lambda}
\def\F{\mathcal{F}}
\def\D{\Delta}
\def\d{\delta}
\def\s{\sigma}
\def\ti{\tilde}
\def\er{\mathrm{erfi}}
\def\integ{\int_{-\infty}^{\infty}}
\def\a{\alpha}
\usepackage[dvips]{color}
\usepackage{here}
\usepackage{amsmath,amssymb}
\slugcomment{}
\shortauthors{Hirano et al.}
\shorttitle{Rossiter-McLaughlin Effect for Transiting Exoplanets}
\begin{document}
\title{Analytic Description of the Rossiter-McLaughlin Effect 
for Transiting Exoplanets: Cross-Correlation Method and Comparison 
with Simulated Data}
\author{
Teruyuki Hirano\altaffilmark{1}, 
 Yasushi Suto\altaffilmark{1,2,3},
 Atsushi Taruya\altaffilmark{1,2,4}, 
 Norio Narita\altaffilmark{5},
 Bun'ei Sato\altaffilmark{6},
 John Asher Johnson\altaffilmark{7}, and
 Joshua N.\ Winn\altaffilmark{8}
} 
\altaffiltext{1}{Department of Physics, The University of Tokyo, 
Tokyo 113-0033, Japan}
\altaffiltext{2}{Research Center for the Early Universe, School of Science, 
The University of Tokyo, Tokyo 113-0033, Japan
}
\altaffiltext{3}{Department of Astrophysical Sciences, 
Princeton University, Princeton, NJ 08544}
\altaffiltext{4}
{Institute for the Physics and Mathematics of the Universe (IPMU), 
The University of Tokyo, Chiba 277-8582, Japan}
\altaffiltext{5}{National Astronomical Observatory of Japan, 
2-21-1 Osawa, Mitaka, 
Tokyo, 181-8588, Japan}
\altaffiltext{6}{Global Edge Institute, Tokyo Institute of Technology, 
2-21-1 Ookayama, Meguro, Tokyo 152-8550, Japan}
\altaffiltext{7}{
Department of Astrophysics, The California Institute of Technology,
MC 249-17
Pasadena, CA 91125
}
\altaffiltext{8}{Department of Physics, and Kavli Institute 
for Astrophysics and Space Research, Massachusetts Institute of Technology,
 Cambridge, MA 02139}
\email{hirano@utap.phys.s.u-tokyo.ac.jp}
\begin{abstract}
  We obtain analytical expressions for the velocity anomaly due to the
  Rossiter-McLaughlin effect, for the case when the anomalous radial
  velocity is obtained by cross-correlation with a stellar template
  spectrum.  In the limit of vanishing width of the stellar absorption
  lines, our result reduces to the formula derived by
  \citet{Ohta2005}, which is based on the first moment of distorted
  stellar lines. Our new formula contains a term dependent on the
  stellar linewidth, which becomes important when rotational line
  broadening is appreciable.  We generate mock transit spectra for 
  four existing exoplanetary systems (HD17156, TrES-2,
 TrES-4, and HD209458) following the procedure of
  \citet{Winn2005}, and find that the new formula is in better
  agreement with the velocity anomaly extracted from the mock
  data. Thus, our result provides a more reliable analytical
  description of the velocity anomaly due to the Rossiter-McLaughlin
  effect, and explains the previously observed dependence of the
  velocity anomaly on the stellar rotation velocity.
\end{abstract}

\section{Introduction \label{s:intro}}

Among approximately 400 extrasolar planets discovered as of September
2009, there are more than 60 transiting planets. The transiting
planets provide important information, such as radii and atmospheric
signatures, which are unavailable from radial-velocity data
alone. Another advantage of transiting exoplanets is that one can
measure the angle between the stellar rotation axis and the orbital
axis of the exoplanet projected onto the sky (conventionally denoted
by $\la$) through the Rossiter-McLaughlin (RM) effect. The RM effect
generates a radial velocity anomaly during a transit, due to the
asymmetric line profiles that result from the partial occultation of
the rotating stellar disk \citep{Rossiter1924, McLaughlin1924}. By
carefully investigating velocity anomalies during transits, one can
determine the trajectory of planets on the stellar disk and estimate
$\la$ precisely \citep[e.g.,][]{Queloz2000,Ohta2005,Winn2005}.

This angle is especially important in understanding the basic physical
processes of planetary formation and subsequent orbital migration.
According to recent planetary formation theories, gaseous planets
orbiting within 0.1 AU of the parent star (hot-Jupiters) are supposed
to have formed a few AU away from the star, and subsequently to have
migrated inward.  While standard migration mechanisms keep $\la\sim
0^\circ$, scenarios such as planet-planet scattering may change $\la$
significantly \citep{Lin1996,Chatterjee2008,Nagasawa2008}.  The Kozai
mechanism may also lead to a highly inclined orbit \citep{Wu2007}.
Thus, the observed distribution of the angles provides an
observational clue to distinguish between different planet formation
and migration theories.

\citet{Ohta2005} derived an analytic formula (hereafter, the OTS
formula) for velocity anomalies due to the RM effect, by computing the
first moment (intensity-weighted mean wavelength) of distorted
absorption lines perturbatively.  (Previously, \citet{Kopal1942} and
others used the first-moment method for eclipsing binary stars to
estimate the velocity anomalies.)  The OTS formula proved to be useful
in understanding the parameter dependence of the velocity anomaly and
also in forecasting the error budget of the parameter estimate, in
particular of the spin-orbit misalignment angle $\lambda$. Indeed
their work inspired \citet{Winn2005} to revisit the estimate of
$\lambda$ precisely for the first discovered transiting planetary
system HD~209458.  For that purpose, \citet{Winn2005} generated
in-transit stellar spectra of HD~209458, put them into the Keck
analysis routine of radial velocities, and found that the OTS formula
systematically underpredicts the velocity anomaly by
10\%. Intriguingly, however, \citet{Johnson2008} and \citet{Winn2008},
found the OTS formula to provide an adequate description of the
simulated results for the cases of HAT-P-1 and TrES-2.

We noticed that the systematic difference between the OTS formula and
the simulation may be sensitive to the stellar spin rotation velocity
$v\sin i_\star$ ($i_\star$ denotes the inclination angle of the
stellar spin axis), which is $4.70 \pm 0.16~ {\rm km} ~{\rm s}^{-1}$,
$3.75 \pm 0.58~ {\rm km}~ {\rm s}^{-1}$, and $1.0 \pm 0.6~ {\rm km}
~{\rm s}^{-1}$, for the HD~209458, HAT-P-1, and TrES-2 systems,
respectively. 
In addition, the actual radial velocity measurement
algorithm does not adopt the moment method, strictly speaking.  For
instance, the analysis routines used by the HARPS (the High Accuracy
Radial velocity Planet Searcher) and SOPHIE (Spectrographe pour
l'Observation des Ph$\mathrm{\acute{e}}$nom$\mathrm{\grave{e}}$nes des
Int$\mathrm{\acute{e}}$rieurs stellaires et des
Exoplan$\mathrm{\grave{e}}$tes) teams involve the cross-correlation of
the observed spectrum with a template spectrum, and the velocity
anomaly is estimated from the peak of the cross-correlation function.
The analysis routine which uses the iodine cell technique, with the
Subaru HDS (the High Dispersion Spectrograph) or Keck HIRES (the High
Resolution Echelle Spectrometer), involve a fit to the observed
spectra, based on a template spectrum and the known spectrum of the
iodine cell.

For these reasons, in this paper we revisit the analytic approach to
the RM effect using the cross-correlation method.  Our analytic
approach to this problem is complementary to the recent numerical
approach by \citet{Triaud2009}, and provides an analytic framework for
understanding their results.  Even though the cross-correlation method
is not directly applicable to the analysis of the iodine cell
technique, we find that our result seems to capture the main
qualitative features of the numerical results based on that technique;
specifically, our formula includes a term dependent on the stellar
spin velocity that does not show up in the moment method but was
empirically found by \citet{Winn2005}.

The rest of the paper is organized as follows. Section 2 describes our
analytical modeling of stellar absorption lines distorted by the
occultation due to a transiting planet (\S 2.1) and derives the RM
velocity anomaly on the basis of the moment method (\S 2.2) and the
cross-correlation method (\S 2.3).  In order to check the
reliability of the new analytic formula, we generate simulated spectra
during transits and put them into the Subaru analysis routine of
radial velocities in \S 3 as \citet{Winn2005} did for the Keck
routine. Specifically we consider HD~17156 ($v\sin i_\star =4.2~ {\rm
  km}~ {\rm sec}^{-1}$), TrES-2 ($v\sin i_\star =1.0 ~{\rm km}~
{\rm sec}^{-1}$), TrES-4 ($v\sin i_\star =8.5 ~{\rm km}~
{\rm sec}^{-1}$),
and HD209458 ($v\sin i_\star =4.5 ~{\rm km}~
{\rm sec}^{-1}$) systems, and find that our new formula reproduces the
simulated data better than the OTS formula.  Finally, section 4 is
devoted to a summary and further discussion of the present paper.

\section{Analytic Model for the Velocity Anomaly \label{s:sec2}}

In this section, we analytically derive the velocity
anomaly due to the RM effect based on the two different methods.  The
final result for the moment method is equation (\ref{Dv}). For the
cross-correlation method, the key result is equation (\ref{gaussianDv})
when the stellar line profile and the rotational broadening kernel are
approximated by equations (\ref{gaus}) and (\ref{eq:RapproxG}),
respectively.  Readers who are not interested in the mathematical
details of the derivations may simply skip this section and directly
move to \S 3.

\subsection{Velocity Anomaly During Transit}

\begin{figure}[h]
\begin{center}
\includegraphics[width=8.5cm]{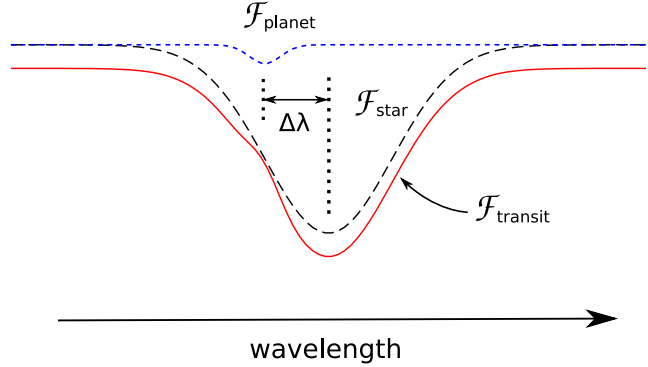}
\end{center}
\caption{ Schematic absorption line profiles of a transiting planetary
system: $\mathcal{F}_{\mathrm{star}}(\lambda)$ is a symmetric line
profile with respect to the central wavelength of each line $\lambda_0$
over the entire stellar disk outside the transit, while
$\mathcal{F}_{\mathrm{planet}}(\lambda-\Delta \lambda)$ is a symmetric
line profile for the portion occulted by a planet that is shifted by a
wavelength of $\Delta\lambda$ with respect to $\lambda_0$.  The
resulting distorted stellar line profile in transit is given by
$\mathcal{F}_{\mathrm{transit}}(\lambda) =
\mathcal{F}_{\mathrm{star}}(\lambda) -
\mathcal{F}_{\mathrm{planet}}(\lambda-\Delta\lambda)$.
\label{lineprofile} }
\end{figure}

The stellar absorption line profile distorted by the RM effect is
schematically illustrated in Figure \ref{lineprofile}.  During a
transit, a portion on the stellar disk is occulted by a planet, and the
flux contribution from the portion occulted by the planet,
$\mathcal{F}_{\mathrm{planet}}(\lambda-\Delta\lambda)$ is subtracted
from the stellar absorption line,
$\mathcal{F}_{\mathrm{star}}(\lambda)$, yielding a distorted absorption
line profile in transit, $\mathcal{F}_{\mathrm{transit}}(\lambda)$.  The
center of $\mathcal{F}_{\mathrm{planet}}$ is Doppler-shifted by $\Delta
\la$ relative to the central wavelength $\lambda_0$ of
$\mathcal{F}_{\mathrm{star}}$ since the portion occulted by the planet
has a stellar rotation velocity:
\begin{equation}
\label{eq:vp-deltalambda}
 v_p = c \frac{\Delta\la}{\lambda_0}
\end{equation}
along the line of sight of an observer.  Because of
the distortion, the absorption line profile becomes asymmetric,
producing an apparent velocity anomaly when the radial velocity is
estimated using one of the standard techniques.

\begin{figure}[h]
\begin{center}
\includegraphics[width=13cm]{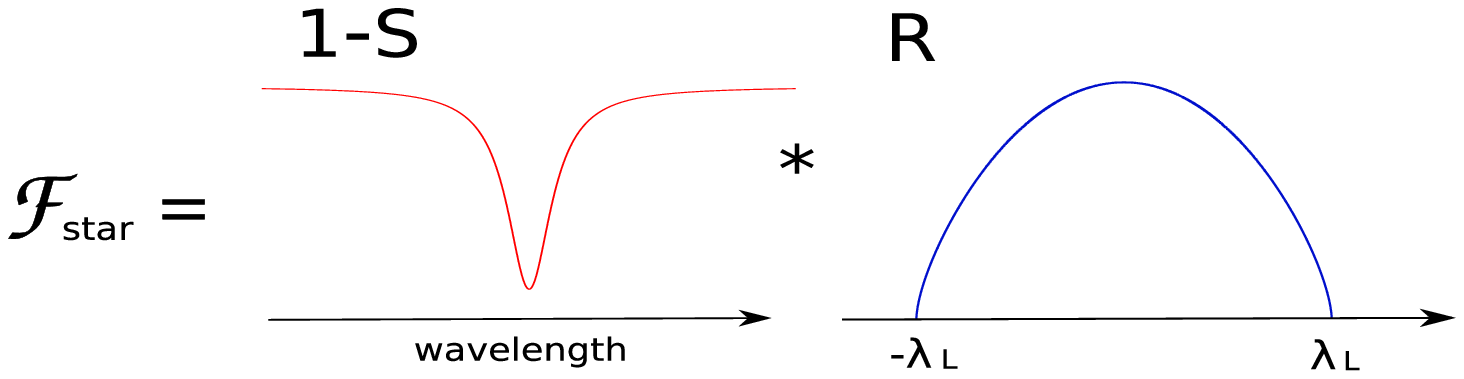}
\end{center}
\caption{Stellar absorption line profile $\mathcal{F}_{\mathrm{star}}$,
 which is a convolution of the intrinsic stellar line profile (including
 the natural broadening), $S(\lambda)$ 
and the stellar rotational kernel $R(\lambda;\lambda_L)$. 
}\label{stellarflux}
\end{figure}

Let us first write a stellar absorption line profile as a convolution of
an intrinsic line profile $S(\lambda)$ and a stellar rotation kernel
$R(\lambda;\la_L)$ as
\begin{eqnarray}
\label{eq:fstar}
\mathcal{F}_\mathrm{star}(\lambda)=1-S(\lambda)*R(\lambda;\la_L),
\end{eqnarray}
where the symbol * indicates a convolution, and the continuum level is
normalized to unity (Fig. \ref{stellarflux}).  In this section, we
implicitly focus on a particular single line of a central wavelength
$\lambda_0$, and locate the center of $S(\la)$ at
$\lambda_0$.  Then we normalize $S(\la)$ and $R(\la; \la_L)$ so that
\begin{eqnarray}
\label{eq:sr-normalization}
\integ S(\lambda) d\lambda =1, \qquad
\integ R(\lambda;\lambda_L) d\lambda =1.
\end{eqnarray}

A Voigt profile (convolution of a Gaussian and a Lorentzian; see
eq.~[\ref{eq:SGandL}] below) is often used to approximate $S(\lambda)$,
but a Gaussian profile also provides a reasonable approximation.  We
will mainly be concerned with the Gaussian approximation in \S~2.3.

The function $R(\la;\la_L)$ indicates a rotational broadening kernel and
broadens the intrinsic stellar line at $\lambda_0$ over the range of
$\lambda_0-\la_L$ and $\lambda_0 + \la_L$, where $\la_L/\lambda_0 = v\sin
i_\star/c$. Its expression for a rigid rotating stellar disk can be
computed once the limb darkening law is specified
\citep[see, e.g.,][]{Gray2005}.  In this paper, we consider the quadratic
limb darkening law in which the intensity of the stellar disk at a
position $(x, y)$ relative to the center of the star is expressed as
\begin{eqnarray}
\label{limbdark}
\frac{I(x,y)}{I_c}
=1-u_1(1-\cos\theta)-u_2(1-\cos\theta)^2,~~\cos \theta 
=\sqrt{1-\frac{x^2+y^2}{R_s^2}}.
\end{eqnarray}
In the above expression, $\theta$ is the angle between the line-of-sight
and the normal vector to the local stellar surface, $R_S$ is the stellar
radius, the coefficients $u_1$ and $u_2$ are the limb-darkening
parameters, and $I_c$ indicates the intensity at the center of the
stellar disk.

Then we obtain
\begin{eqnarray}
\label{rotation}
R(\la; \la_L) = 
c_1\sqrt{1-\left(\frac{\la}{\la_L}\right)^2}
+c_2\left\{1-\left(\frac{\la}{\la_L}\right)^2\right\}
+c_3\left\{1-\left(\frac{\la}{\la_L}\right)^2\right\}^{3/2},
\end{eqnarray}
where the coefficients $c_1$, $c_2$, and $c_3$  are given as
\begin{eqnarray}
c_1&=&\frac{2(1-u_1-u_2)}{\pi\la_L(1-u_1/3-u_2/6)},\label{c_1}\\
c_2&=&\frac{u_1+2u_2}{2\la_L(1-u_1/3-u_2/6)},\label{c_2}\\
c_3&=&-\frac{4u_2}{3\pi\la_L(1-u_1/3-u_2/6)}.\label{c_3}
\end{eqnarray}

Adopting equation (\ref{eq:fstar}) as a line profile, the line distortion
due to the RM effect is described as follows. In the transiting
planetary system, the portion occulted by the planet is sufficiently
small. Thus the effect of the rotational broadening within the portion
is safely negligible, and one can write
\begin{eqnarray}
\label{eq:fplanet}
\F_{\mathrm{planet}}(\la-\D\la)=f\left\{1-S(\lambda-\D\la)\right\},
\end{eqnarray}
where $f$ is the ratio of the flux from the occulted part of the stellar
disk to the total flux.  

In general, $\D\la= v_p\lambda_0/c$ is a function of the position of
the planet and the contribution of wavelength shift is determined by
integrating over the entire occulted portion of stellar disk.  The
precise definition of $v_p$ in terms of the stellar rotation velocity
and the position of planet is given by equation (\ref{subp}), which is
simplified as equation (\ref{subp3}) for rigid rotation.

Then, the line profile during transit is given as
\begin{eqnarray}
\label{eq:ftransit}
\F_{\mathrm{transit}}(\la)
&=&\F_{\mathrm{star}}(\la)-\F_{\mathrm{planet}}(\la-\D\la)\nonumber\\
&=&\left\{1-S(\lambda)*R(\lambda;\la_L)\right\}
-f\left\{1-S(\la-\D\la)\right\}.
\end{eqnarray}
Because the continuum level $1-f$ in equation (\ref{eq:ftransit}) does
not change the evaluation of the velocity anomaly, in what follows we
simply use the expression
\begin{eqnarray}
\label{ftra}
\F_{\mathrm{transit}}(\la)
=-S(\lambda)*R(\lambda;\la_L)+fS(\la-\D\la)
\end{eqnarray}
without losing generality.

\subsection{Estimate Based on the Moment Method}

Let us consider first the moment method, in which we estimate the
wavelength shift of the center of a distorted line during a transit as
\citep[e.g.][]{Ohta2005}:
\begin{eqnarray}
\label{moment}
\delta_{\rm RM} \equiv 
\frac{\displaystyle\int_{-\infty}^{\infty}
\lambda \F_\mathrm{transit}(\lambda)d\lambda}
{\displaystyle\int_{-\infty}^{\infty}
\F_\mathrm{transit}(\lambda)d\lambda}.
\end{eqnarray}

Substituting equation (\ref{eq:ftransit}) into
$\mathcal{F}_\mathrm{transit}(\la)$ in equation (\ref{moment}), the
denominator is
\begin{eqnarray}
\label{-1+f}
\integ\F_{\mathrm{transit}}(\la)d\lambda
&=&-\integ d\la^\prime R(\la^\prime;\la_L) \integ d\la S(\la- \la^\prime)
+f\integ d\la S(\la-\D\la)\nonumber\\
&=&-1+f.
\end{eqnarray}
The numerator of equation (\ref{moment}) reduces to
\begin{eqnarray}
\label{fdla}
\integ\la\F_{\mathrm{transit}}(\la)d\la
&=&-\integ d\la^\prime R(\la^\prime;\la_L)\integ \la S(\la-\la^\prime)d\la
+f\integ \la S(\la-\D\la)d\la\nonumber\\
&=&-\integ  R(\la^\prime;\la_L)d\la^\prime 
\integ (\la-\la^\prime) S(\la-\la^\prime)d\la\nonumber\\
&&-\integ  \la^\prime R(\la^\prime;\la_L)d\la^\prime 
\integ S(\la-\la^\prime)d\la\nonumber\\
&&+f\integ (\la-\D\la)S(\la-\D\la)d\la+f\integ \D\la S(\la-\D\la)d\la\nonumber\\
&=&f\integ \D\la S(\la-\D\la)d\la\nonumber\\
&=&f\D\la ,
\end{eqnarray}
and therefore we obtain
\begin{eqnarray}
\label{eq:deltarm-moment}
\d_\mathrm{RM} = -\frac{f}{1-f}\D\la.
\end{eqnarray}

Using equation (\ref{eq:vp-deltalambda}), one can rewrite
equation (\ref{eq:deltarm-moment}) in terms of the velocity anomaly as
\begin{eqnarray}
\label{Dv}
\D v \equiv c \frac{\delta_{\rm RM}}{\lambda_0}
=-\frac{f}{1-f}v_p .
\end{eqnarray}

When we explicitly write the flux ratio $f$ and the subplanet velocity
$v_p$ as a function of the planet position, we reproduce the OTS formula
(see Appendix \ref{s:flux} and \ref{s:ingress} below).  In particular,
the above expression coincides with equation (25) of OTS which is the
formula for complete transiting case without limb-darkening, if we
simply replace the quantities of $f$ and $v_p$ with $\gamma^2$ and
$\Omega_sx\sin i_\star$.  

Note that equations (\ref{-1+f}) to (\ref{eq:deltarm-moment}) are
derived from the normalization condition (\ref{eq:sr-normalization})
of $S(\lambda)$ and $R(\lambda;\lambda_L)$ and the fact that they are
even functions of $\lambda$, and do not depend on their specific
functional forms. This implies that the velocity anomaly derived from
the moment method depends on neither the profile nor the width of the
absorption line. Thus the results are independent of the stellar
rotation velocity.

%
\subsection{Estimate Based on the Cross-Correlation Method}

\subsubsection{Formulation for the Voigt profile}

Consider next the estimate of the velocity anomaly based on the
cross-correlation method. In this case, the wavelength shift of central
line $\d_\mathrm{RM}$ is obtained by maximizing the cross-correlation
function:
\begin{eqnarray}
\label{deriva}
&& \frac{dC(x)}{dx}\Big|_{x=\delta_\mathrm{RM}} = 0, \\
\label{Cross}
&& C(x) = \integ \F_{\mathrm{star}}(\la-x) \F_\mathrm{transit}(\la)d\la.
\end{eqnarray}
To proceed further, we have to specify the functional form of $S(\la)$
and $R(\la, \la_L)$.

Here we adopt the Voigt profile for the stellar line:
\begin{eqnarray}
\label{eq:SGandL}
S(\la) &=& G_{\rm S}(\la;\beta)*L(\la;\gamma), \\
\label{gaus}
G_{\rm S}(\la; \beta)&=&\frac{1}{\beta\sqrt{\pi}}e^{- \la^2/\beta^2}, \\
\label{lore}
L(\la; \gamma_\star)&=&\frac{1}{\pi}\frac{\gamma_\star}{\la^2+\gamma_\star^2}.
\end{eqnarray}
In the above expressions, $\beta$ in the Gaussian kernel characterizes
the thermal and microturbulent broadening, while $\gamma_\star$ in the
Lorentzian kernel represents the line broadening due to natural
broadening and the Stark effect.

While we derive equation (\ref{rotation}) for the rotational kernel
$R(\la;\la_L)$ for the quadratic limb-darkening law, the expression is
not analytically tractable. Thus we approximate it by a Gaussian:
\begin{eqnarray}
\label{eq:RapproxG}
R(\la;\la_L)
\approx G_{\rm R}(\la;\sigma)
\equiv \frac{1}{\sigma\sqrt{\pi}}e^{- \la^2/\sigma^2} ,
\end{eqnarray}
where the relation between $\sigma$ and $\la_L$ is obtained by
least-square fitting of Gaussian to equation (\ref{rotation}).  Their
explicit relation is given in Appendix \ref{s:leastsquare}.

Under this Gaussian approximation for the rotation kernel, we have
\begin{eqnarray}
S(\lambda)*R(\la;\la_L)
\approx G_{\rm S}(\la;\beta)*L(\la;\gamma_\star)*G_{\rm R}(\la;\sigma)
\equiv G(\la;\beta_\star)*L(\la;\gamma_\star) ,
\end{eqnarray}
where
\begin{eqnarray}
\beta_\star^2 \equiv \beta^2 +\sigma^2 .
\end{eqnarray}
This significantly simplifies the analytic computation, and
we now have 
\begin{eqnarray}
\label{F1}
\F_\mathrm{transit}(\la)
&=& -\integ G(\lambda^\prime;\beta_\star)
 L(\la-\la^\prime;\gamma_\star)d\la^\prime \cr
&&~~ +f \integ
 G(\la^\prime;\beta_p)L(\la-\D\la-\la^\prime;\gamma_p)d\la^\prime\\
\label{F2}
\F_\mathrm{star}(\la) &=&
- \integ G(\lambda^\prime;\beta_\star)
L(\la-\la^\prime;\gamma_\star)d\la^\prime,
\end{eqnarray}
where $\beta_p$ and $\gamma_p$ indicate the narrow line originated from
the portion of the planet, and thus we simply set $\beta_p=\beta$.

Substituting equations (\ref{F1}) and (\ref{F2}) into equations (\ref{deriva}) 
and (\ref{Cross}), we obtain the following algebraic equation for
$\d_\mathrm{RM}$:
\begin{eqnarray}
\label{exact}
&\displaystyle\frac{1}{b^{3/2}}e^{-(a^2-c^2)/4b}
~\mathrm{Re}\left\{(a-ic)e^{iac/2b}
\left[1+\mathrm{erfi}\left(\frac{a-ic}{2\sqrt{b}}\right)\right]\right\} \cr
&=\displaystyle \frac{f}{B^{3/2}}e^{-(A^2-C^2)/4B}
~\mathrm{Re}\left\{(A-iC)e^{iAC/2B}
\left[1+\mathrm{erfi}\left(\frac{A-iC}{2\sqrt{B}}\right)\right]\right\},
\end{eqnarray}
where
\begin{eqnarray}
&a=2\pi \d_\mathrm{RM}, ~
b=2\pi^2\beta_\star^2, ~
c=4\pi\gamma_\star\label{small}\\
&A=2\pi(\d_\mathrm{RM}-\D\la),~
B=\pi^2(\beta_\star^2+\beta_p^2),~
C=2\pi(\gamma_\star+\gamma_p).\label{large},
\end{eqnarray}
and $\mathrm{erfi}(z)$ denotes the imaginary error function
defined in terms of the error function erf$(x)$:
\begin{eqnarray}
\er(z) \equiv -i ~\mathrm{erf}(iz)
=\frac{2}{i\sqrt{\pi}}\int_0^{iz}e^{-t^2}dt.
\end{eqnarray}
The derivation of equation (\ref{exact}) is summarized in Appendix
\ref{s:derivation}.

\subsubsection{Analytic formula for the Gaussian profile}

Since the Voigt profile is widely used for stellar absorption lines,
equation (\ref{exact}) is fairly general. While it cannot be solved
exactly for $\delta_{\rm RM}$, we obtain an approximate perturbation
solution for $\delta_{\rm RM} \ll \Delta\lambda$. In particular, we
present the result for the Gaussian profile ($\gamma_\star=\gamma_p=0$)
in this subsection, which will be compared with simulated data analysis
in \S 3.  The other solutions for the Lorentzian profile
($\beta_\star=0$) and the Gaussian profile with a small Lorentzian wing
($\gamma_\star \ll \beta_\star$) are presented in Appendix
\ref{s:lorentziansection} and \ref{s:perturbation}, respectively.

In the Gaussian profile assumed here, the parameters $c$ and $C$ in
equations (\ref{small}) and (\ref{large}) vanish. Then equation
(\ref{exact}) reduces to
\begin{eqnarray}
\displaystyle\frac{a}{b^{3/2}}e^{-a/4b}=f\frac{A}{B^{3/2}}e^{-A/4B},
\end{eqnarray}
or equivalently
\begin{eqnarray}
\label{gaussexact}
f=\left(\frac{\beta_\star^2+\beta_p^2}{2\beta_\star^2}\right)^{3/2}
\frac{\d_\mathrm{RM}}{\d_\mathrm{RM}-\D\la}
\exp\left\{-\frac{\d_\mathrm{RM}^2}{2\beta_\star^2}
+\frac{(\d_\mathrm{RM}-\D\la )^2}{\beta_\star^2+\beta_p^2}\right\}.
\end{eqnarray}

Note that $f$ is approximately the square of the planet-to-star radius
ratio, and is of order $10^{-2}$ for giant planets. Thus, equation
(\ref{eq:deltarm-moment}) implies that $|\delta_{\rm RM}/\Delta
\lambda| \approx f \ll 1$, and we can expand the exponent in equation
(\ref{gaussexact}) up to the linear order of $\delta_{\rm RM}$:
\begin{eqnarray}
-\frac{\d_\mathrm{RM}^2}{2\beta_\star^2}
+\frac{(\d_\mathrm{RM}-\D\la )^2}{\beta_\star^2+\beta_p^2}
\approx \frac{\D\la^2}{\beta_\star^2+\beta_p^2}
\left(1-2\frac{\d_\mathrm{RM}}{\D\la}\right)
\equiv\kappa\left(1-2\frac{\d_\mathrm{RM}}{\D\la}\right).
\end{eqnarray}
In general we consider the case where $\Delta\lambda$ is well within the
width of the stellar broadening, and therefore $\kappa \ll 1$.
Therefore we expand equation (\ref{gaussexact}) up to the linear order
in $\d_\mathrm{RM}/\D \la$ as
\begin{eqnarray}
f \approx - \left(\frac{\beta_\star^2+\beta_p^2}{2\beta_\star^2}\right)^{3/2}
\frac{\d_\mathrm{RM}}{\D\la}(1+\kappa),
\end{eqnarray}
and finally obtain
\begin{eqnarray}
\d_\mathrm{RM}
\approx 
-\left(\frac{2\beta_\star^2}{\beta_\star^2+\beta_p^2}\right)^{3/2}
f\D\la\left(1-\frac{\D\la^2}{\beta_\star^2+\beta_p^2}\right).
\end{eqnarray} 
As before, the above equation can be rewritten as
\begin{eqnarray}
\label{gaussianDv0}
\D v \approx
 - \left(\frac{2\beta_\star^2}{\beta_\star^2+\beta_p^2}\right)^{3/2}
fv_p\left\{1-\frac{\la_0^2}{c^2(\beta_\star^2+\beta_p^2 )}v_p^2\right\}
\end{eqnarray}
or
\begin{eqnarray}
\label{gaussianDv}
\D v \approx
-\left\{\frac{2(\beta^2+\sigma^2)}{2\beta^2+\sigma^2}\right\}^{3/2}
fv_p\left\{1-\frac{\la_0^2}{c^2(2\beta^2+\sigma^2 )}v_p^2\right\}, 
\end{eqnarray}
where we replace $\beta_p^2$ with $\beta^2$ (line width without
rotational broadening), and $\beta_\star^2$ with $\beta^2+\sigma^2$ in
which $\sigma$ indicates the stellar rotation width.

Equation (\ref{gaussianDv}) is the key result of the present paper which
describes the RM velocity anomaly under the Gaussian approximation for
the intrinsic line profile and the stellar rotation kernel.  In marked
contrast to equation (\ref{Dv}) based on the moment method, equation
(\ref{gaussianDv}) indicates that the radial velocity anomaly from the
cross-correlation method depends both on the width of the line profile
($\beta$) and on the rotation velocity ($\sigma$).  It recovers the
OTS formula only when $\sigma=0$ and the second term in the parenthesis
is negligibly small.

Indeed the presence of the cubic term in $v_p^3$ is consistent with
the empirical finding from the simulated data \citep{Winn2005}.  Note
that a quintic (5th order term) was needed to match the simulations
for the rapidly rotating star HAT-P-2 \citep{Winn2007}, which would
emerge naturally in our formulation by expanding the next-order term
in $\kappa$.  We should note here that in many cases the estimate of
the spin-orbit misalignment angle $\lambda$ is not very sensitive to
the choice of the moment method or the cross-correlation method
discussed in this section; when $\lambda$ is small and the transit
impact parameter is not too close to zero, then $\lambda$ is mainly
determined by the epoch of the photometric central transit when $\D
v$, and thus $v_p$, vanish.

\section{Comparison with Simulated Results 
Based on the Subaru Analysis Routine}

So far, we have focused on the RM velocity anomaly estimated from
single line profiles characterized by the Voigt function.  The
practical data analysis, however, is much more complicated; the
wavelength shift $\Delta\lambda$ is estimated statistically from many
different lines, and each line is not always approximated by a single
Voigt profile.  In addition, Subaru and Keck analysis routines are not
related in a straightforward way to the cross-correlation algorithm.
This is why we create mock spectrum data during transit, compute
radial velocity anomalies using the Subaru analysis routine, and
compare the output results with our analytic formulae, in particular
equation (\ref{gaussianDv}).

\subsection{Subaru Analysis Routine}

In order to check the validity of our analytic formula, we create mock
spectra in transit, and reduce the data with the analysis routine
\citep{Sato2002} for the Subaru HDS radial velocity measurement with
I$_2$ cell. The routine has been extensively used for the previous
observation of the RM effect \citep[e.g.][]{Narita2007,Narita2008}.

The routine finds the best-fit value of the wavelength shift
$\delta\lambda$ by modeling an observed spectrum $I_\mathrm{obs}(\la)$
with the given transmission function of the I$_2$
cell, $A(\la)$, as
\begin{eqnarray}
\label{model}
I_\mathrm{obs}(\la) = k[A(\la) S(\la-\d \la)] *\mathrm{IP},
\end{eqnarray}
where $k$ is the normalization factor, $S(\la-\d \la)$ is the intrinsic
stellar spectrum shifted by $\d\la$, and IP denotes the instrumental
profile.  \citet{Sato2002} use the Lick-Hamilton I$_2$ cell spectrum for
$A(\la)$ with high spectral resolution ($R\sim 400,000$).  The peculiar
velocity and the Keplerian motion of the star as well as
the RM velocity anomaly contribute to $\d\la$, and thus the former two
have to be subtracted in order to extract $\delta_{\rm RM}$ alone.

Once $A(\la)$ and $S(\la)$ are given, the routine simultaneously finds
the best-fit values for $k$, $\d\la$, parameters characterizing IP
(usually expressed as a linear combination of several Gaussians), and
calibration parameters of wavelength by least-square fit to
$I_\mathrm{obs}(\la)$. In reality, the method of \citet{Sato2002} attempts to
elaborate the intrinsic stellar template spectrum $S(\la)$ as well from
the observed spectrum $I_\mathrm{obs}(\la)$. In practice, it proceeds as
follows:
\begin{enumerate}
\item Choose an initial stellar spectrum $S^0(\la)$ from an
      observational dataset or from a theoretical model.
\item Find the best-fit parameters in the right-hand-side of equation
(\ref{model}), and denote the resultant best-fit spectrum by $I^0(\la)$.
\item Compute the residual:
\begin{eqnarray}
\d S(\la)=I_\mathrm{obs}(\la)-I^0(\la),
\end{eqnarray}
and use
\begin{eqnarray}
S^1(\la) =S^0(\la) +\d S(\la),
\end{eqnarray}
as a revised template for the stellar spectrum.
\item Iterate the above steps until the given accuracy is achieved.
\end{enumerate}

In general, the above method finds the best-fit stellar template
spectrum separately for each $I_\mathrm{obs}(\la)$ observed at different
epochs.  In order to avoid the contamination caused by the I$_2$
absorption lines in using only one spectrum, the set of stellar templates at different
times are averaged and the resulting stellar spectrum $\bar{S}(\la)$ is
used to determine $\d \la$ for each $I_\mathrm{obs}(\la)$.

The Keck HIRES analysis routine by \citet{Butler1996} uses the stellar
template spectrum directly obtained from the observed spectrum
(without the I$_2$ cell) by deconvolving the instrumental profile that
is estimated based on observations of a rapidly-rotating B star using
the I$_2$ cell.  Except for the treatment of the stellar spectrum,
both routines are very similar, and our simulation result below is
applicable to the Keck data analysis.

\subsection{Simulated Spectra}

As mock observational spectra for the Subaru analysis routine
($I_\mathrm{obs}(\la)$ in eq.~[\ref{model}]), we create many simulated
spectra during transits following the procedure of \citet{Winn2005} as
follows:
\begin{enumerate}
\item[(i)] Broaden the NSO solar spectrum \citep{Kurucz1984} so as to
	   the include the effect of stellar rotation\footnote{Although
	   the NSO solar spectrum is a rotationally broadened spectrum
	   whose rotational velocity $v\sin i_\star\simeq 1.85$ km
	   sec$^{-1}$, we additionally broaden the spectrum by 
	   Gaussian broadening kernels.}.  We apply two different
	   broadening kernels as discussed below.
\item[(ii)] Compute the spectra during transit from equation
	    (\ref{eq:ftransit}) using the broadened spectrum described
	    above for $\F_{\mathrm{star}}(\la)$, and the original
	    (unbroadened) spectrum for
	    $\F_{\mathrm{planet}}(\la-\D\la)$.
\item[(iii)] Multiply $\F_{\mathrm{transit}}(\la)$ by $A(\la)$, the
	    transmission function of the I$_2$ cell (eq.~[\ref{model}]).
\item[(iv)] Convolve $\F_{\mathrm{transit}}(\la)A(\la)$ with the
	   instrumental profile, IP, which is approximated by a single
	   Gaussian that best-fits the IP of Subaru/HDS.
	  We call the resultant spectra
	   ``the mock data''.
\item[(v)] Feed the mock spectra into the Subaru analysis routine (\S
	   3.1), and compute $\delta\lambda$.
\end{enumerate}

For the broadening kernel for the stellar rotation in step (i), we first
adopt a Gaussian (\S 3.3.1) so as to compare the analytic formula under
the Gaussian approximation, and then use equation (\ref{rotation}) for
more realistic comparison.

Once the broadening kernel is specified, the remaining parameters are
the flux ratio $f$ and the subplanet velocity $v_p=c\D \la/\la$.  We
repeat the above procedure at grids on $(f, v_p)$ plane. We
select $f=0.00, ~0.004,~ 0.008$, $0.010, ~0.012, ~0.016,~ 0.020$, and 
assign to $v_p$
typically $20$--$30$ different values ranging over several km sec$^{-1}$
for each $f$ in equal interval. Then we compute the velocity anomaly $\D
v$ on those grids of $(f, v_p)$. 

Finally we apply the methodology to four existing planetary systems
(HD17156, TrES-2, TrES-4, and HD209458), and fit the result to the following form
inspired by the Gaussian approximation (eq.~[\ref{gaussianDv}]):
\begin{eqnarray}
\label{quadratic}
\D v=-fv_p(p-qv_p^2) ,
\end{eqnarray}
where $p$ and $q$ are constants fitted for each planetary system, and
depend on the stellar absorption line profile and rotation broadening
among others.

\subsection{Results}

\subsubsection{Ideal case with Gaussian Rotational Kernel}

\begin{figure}[h]
\begin{center}
\includegraphics[width=13cm]{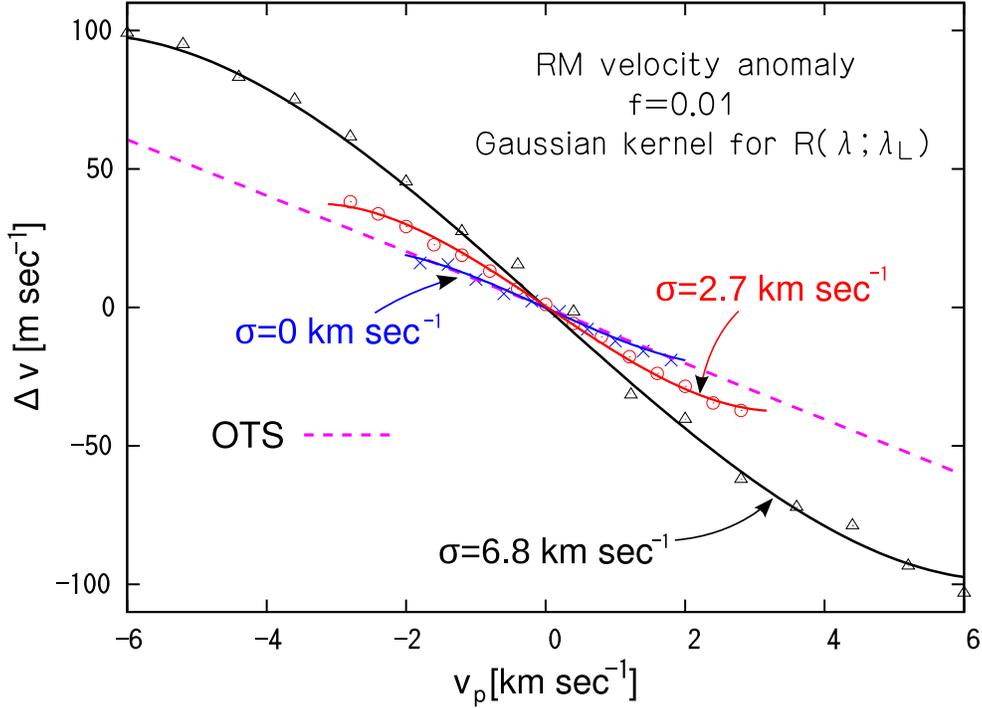} 
\caption{
The RM velocity anomaly against $v_p$ for $f=0.01$.  
{\it symbols}: results extracted from mock data using the Subaru
analysis routine with NSO spectrum for $S(\lambda)$ and the Gaussian
broadening kernel (eq.~[\ref{eq:RapproxG}]). Crosses, circles, and triangles 
correspond to $\sigma=0$, 2.7, and 6.8 km sec$^{-1}$, respectively.  
The number of data points shown is suppressed to make it easier to see.
{\it solid lines}: polynomial fit (eq.~[\ref{quadratic}]) to the mock
results. The OTS formula is plotted in dashed line for comparison
(almost indistinguishable from the $\sigma=0$ line). \label{nf1}
}
\end{center}
\end{figure}

Consider the NSO spectrum additionally broadened with the Gaussian
kernel (eq.~[\ref{eq:RapproxG}]).  We adopt $\sigma=0$, 2.7, and 6.8 km
sec$^{-1}$, \footnote{The dispersion parameter $\sigma$
(or $\beta$) has a unit of length, but we here express them in terms of
radial velocities in order to make it easier to compare the dispersion
parameters with the stellar rotational velocities.}   create mock
spectra during transits for grids on $(f, v_p)$ plane, and determine the
wavelength shift $\D \la$. The velocity anomaly $\Delta
v$ against $v_p$ is plotted in Figure \ref{nf1} for $f=0.01$. Since $\D
v$ is simply proportional to $f$, this plot can be easily scaled for an
arbitrary value of $f$ as long as $f\ll 1$.

Different absorption lines in the NSO spectrum have different line
profiles and are not exactly Gaussian.  If the Gaussian approximation
(eq.~[\ref{gaussianDv}]) were exact, the coefficients in equation
(\ref{quadratic}) for a particular line at wavelength $\lambda_0$ should
be given as
\begin{equation}
p = \left[ 1+ \frac{\sigma^2}{2\beta^2+\sigma^2}\right]^{3/2},
\quad
q = \left[ 1+ \frac{\sigma^2}{2\beta^2+\sigma^2}\right]^{3/2}\frac{\la_0^2}{c^2(2\beta^2+\sigma^2)} .
\end{equation}
Apparently they should be sensitive to each absorption line profile (its
central wavelength $\lambda_0$ and width $\beta$), and it is not clear
that the real data analysis based on many different lines overall
reproduces the polynomial form like equation (\ref{quadratic}).
Nevertheless Figure \ref{nf1} indicates that the simulated results are
indeed fitted very well by equation (\ref{quadratic}). As expected, the
OTS formula derived from the moment method agrees with the simulated
result if the stellar rotational broadening is negligible ($\sigma=0$).

\begin{figure}[h]
\begin{center}
\includegraphics[width=13cm]{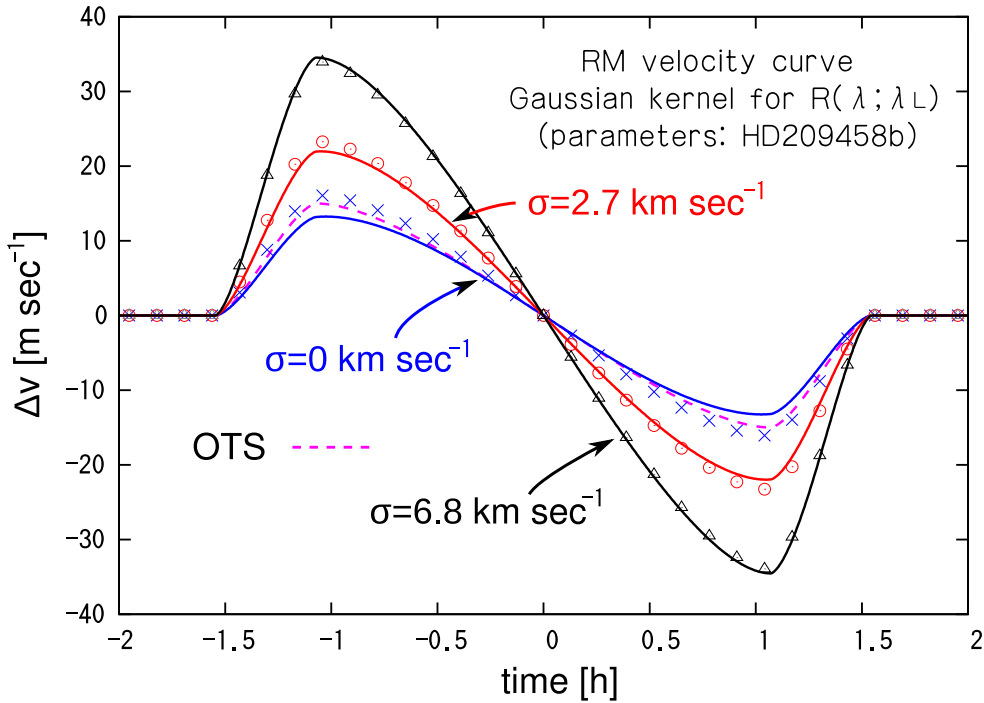} \caption{The RM
velocity anomaly curve for HD209458 but with varying $\sigma$ for the
Gaussian broadening kernel. The central transit time is set to the
origin of the time.  The spin-orbit misalignment angle $\la$ is assumed
to be $0^\circ$.  {\it symbols}: predictions based on the polynomial fit
(eq.~[\ref{quadratic}]) plotted in Figure \ref{nf1}.  Crosses, circles, and triangles
correspond to $\sigma=0$, 2.7, and 6.8 km sec$^{-1}$,
respectively.  {\it solid lines:} analytic formula
(eq.~[\ref{gaussianDv}]) with the Gaussian broadening kernel based on the
cross-correlation method.  The OTS formula is plotted in dashed line for
comparison.  \label{nf2}}
\end{center}
\end{figure}

Given the above encouraging result, we can use equation
(\ref{quadratic}) to compute the velocity anomaly curve.  In order to do
so, we have to compute $f$ and $v_p$ as a function of time, which can be
done using the formulae in Appendix \ref{s:flux} and \ref{s:ingress}
(see also OTS).  Figure \ref{nf2} plots the prediction for HD209458 in
symbols on the basis of the result plotted in Figure \ref{nf1}.  Also
plotted for comparison is the analytic formula (\ref{gaussianDv}) for
the Gaussian kernel adopting $\beta=2.6$ km sec$^{-1}$ as the best-fit
of the original solar spectrum. 
We here adopt $\la_0=5500~\mathrm{\AA}$ 
as a typical wavelength of the radial velocity analysis\footnote{Many absorption 
lines of the iodine cell are located between $5000~\mathrm{\AA}$ and
$6000~\mathrm{\AA}$.}.
Since the actual stellar rotation speed of HD209458
is $v\sin i_\star = 4.70 \pm 0.16~ {\rm km}~ {\rm sec}^{-1}$, $\sigma
\approx 4~{\rm km}~ {\rm sec}^{-1}$ is appropriate, but we consider
$\sigma=0$, 2.7, and 6.8 km sec$^{-1}$ so as to examine the dependence on
$\sigma$.

 Again we stress here that the lines in the NSO spectrum are not
perfectly Gaussian, and the Subaru analysis routine adopts
multi-parameters fitting with the least-squares method, instead of the
cross-correlation method we assumed in the analytic treatment.
Nevertheless the simulated result and the analytic formula
(\ref{gaussianDv}) show good overall agreement, indicating the
practical validity of our analytic approach.


\subsubsection{Application to actual exoplanetary systems}

While we find that the basic features of the simulated are explained by
the Gaussian approximation both of the stellar line profile and the
stellar rotation kernel, it still remains to see the extent to which a
more realistic rotation kernel affects the RM velocity anomaly.  
We examine this point by repeating the simulated analysis for 
HD17156, TrES-2, TrES-4, and HD209458 with the rotation kernel (\ref{rotation}).
\begin{description}
\item[HD17156] This transiting planetary system was discovered
by \citet{Fischer2007}.  We assume the linear limb darkening law with
$u_1=0.6$ and $u_2=0$ and the rotational velocity of $v\sin i_\star=4.2$
km sec$^{-1}$\citep{Narita2009}.\footnote{While Fischer et al. (2007) estimated the
stellar rotational velocity to be $v\sin i_\star=2.6$ km sec$^{-1}$,
Narita et al. (2009) obtained the stellar rotational velocity $v\sin
i_\star=4.2$ km sec$^{-1}$ by the RM analysis.  We adopt the latter
value here.} We perform the same simulation as before but now using the
rotation kernel (\ref{rotation}), and obtain
\begin{eqnarray}
\label{hd17156v}
\D v=-fv_p\left[1.37-0.505\left(\frac{v_p}{v\sin
			    i_\star}\right)^2\right].
\end{eqnarray}
The result is plotted in crosses in Figure \ref{tres4-fig} (a). 
The misalignment angle $\lambda$ is
assumed to be $10^\circ$. We adopt $i_o=87.2^\circ$ as
the orbital inclination of HD17156b.

While the rotation kernel is not Gaussian, it is interesting to see the
difference between our analytical formula and the simulated result.  For
this purpose, we fit equation (\ref{rotation}) by a single Gaussian and
find the best-fit of $\sigma \approx \la_L/1.31$ (see Appendix
\ref{s:leastsquare} for detail).  
The solid line in 
Figure \ref{tres4-fig} (a) indicates the corresponding result from equation
(\ref{gaussianDv}) for $\beta=2.6$ km sec$^{-1}$ and $\sigma=v\sin
i_\star/1.31 \approx 3.2$ km sec$^{-1}$.
\begin{figure}
\begin{minipage}{0.5\hsize}
\begin{center}
\includegraphics[width=8cm]{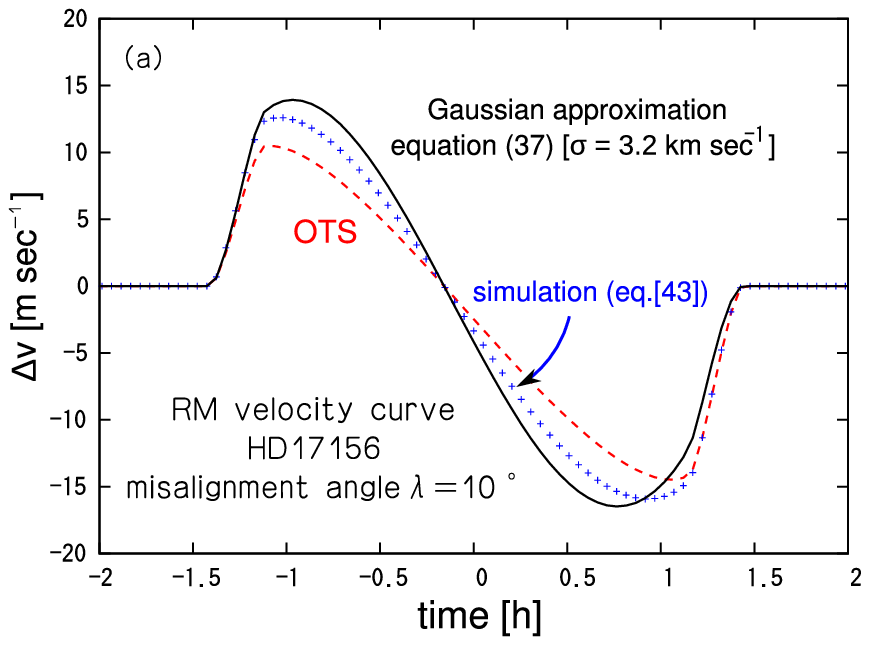}
\end{center}
\end{minipage}
\begin{minipage}{0.5\hsize}
\begin{center}
\includegraphics[width=8cm]{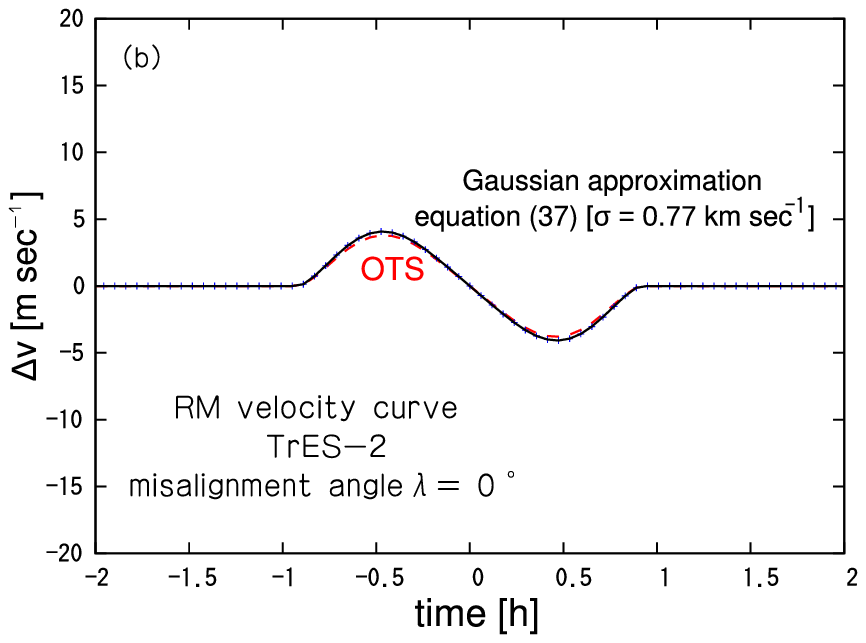}
\end{center}
\end{minipage}
\begin{minipage}{0.5\hsize}
\begin{center}
\includegraphics[width=8cm]{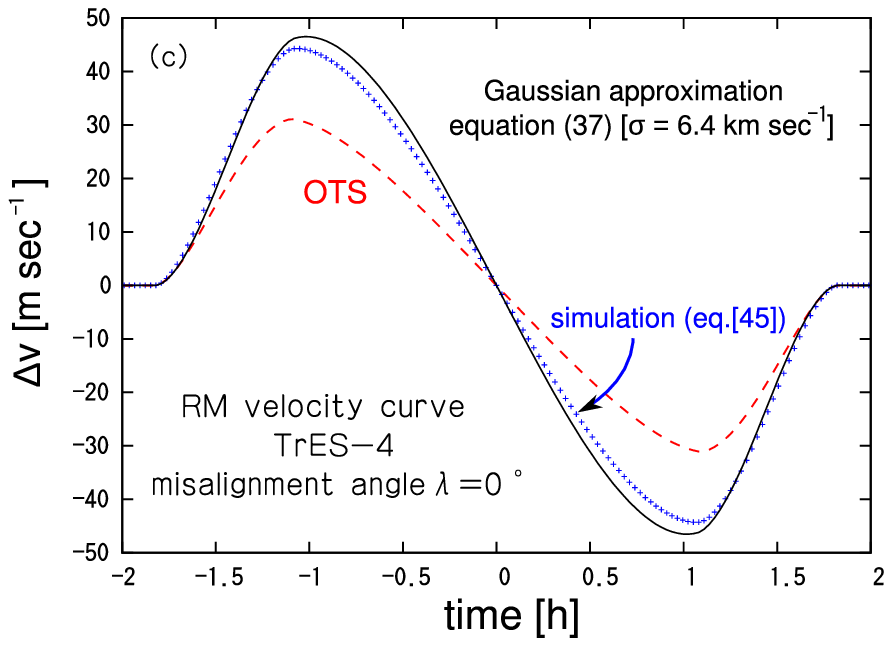}
\end{center}
\end{minipage}
\begin{minipage}{0.5\hsize}
\begin{center}
\includegraphics[width=8cm]{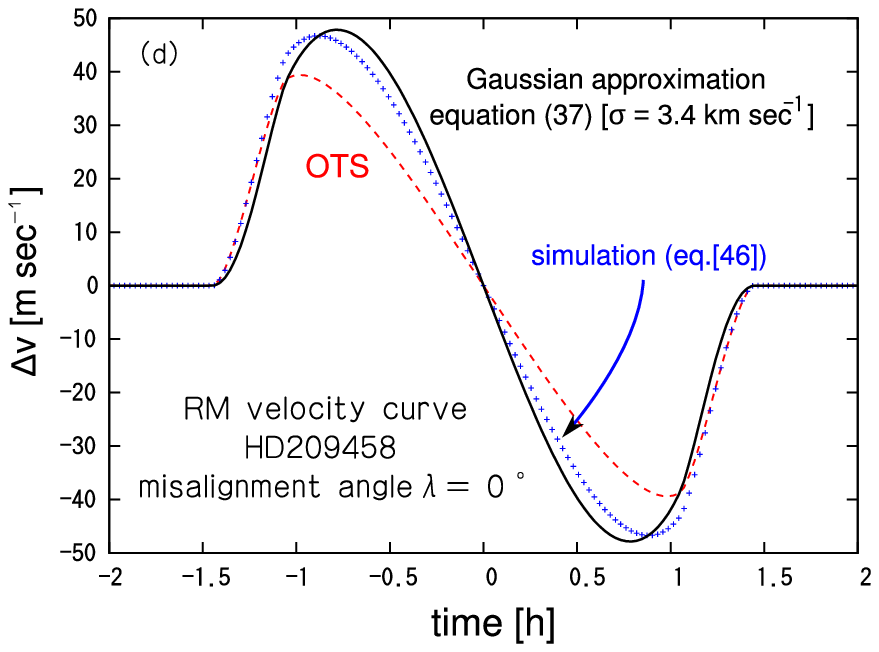}
\end{center}
\end{minipage}
\caption{
The RM
velocity anomaly curves for HD17156 (a), TrES-2 (b), TrES-4 (c), 
and HD209458 (d).  Crosses indicate the simulated result 
using the polynomial fit
(eqs.[\ref{hd17156v}], [\ref{TrES-2}], [\ref{TrES-4}], and [\ref{HD209458}]).  
Solid and dashed lines
correspond to the Gaussian formula (eq.~[\ref{gaussianDv}]) and the OTS
formula, respectively.  The spin-orbit misalignment angle $\lambda$ is
assumed to be 10$^\circ$ for HD17156 and 0$^\circ$ for other systems.  We adopt
$87.2^\circ$, $83.6^\circ$, $82.6^\circ$, and $86.7^\circ$ for the orbital inclination $i_o$ of
HD17156, TrES-2, TrES-4, and HD209458, respectively. 
\label{tres4-fig}
}
\end{figure}
\item[TrES-2] The central star in TrES-2 system has a
small projected rotational velocity. \citet{Winn2008} investigated the
RM effect of this system, and concluded the orbit of TrES-2b is
prograde. They derived the empirical formula for the velocity anomaly
using templates based on observations of similar stars.  Here, we use
a theoretical G0 type template spectrum without rotational broadening
\citep{Coelho2005}, instead of the NSO spectrum, to create the mock
spectra.  We then broaden the theoretical spectrum assuming $v\sin
i_\star=1.0$ km sec$^{-1}$ and the limb-darkening parameters $u_1=0.40$
and $u_2=0.30$. After inputting the mock spectra into the analysis
routine, we obtain the following empirical formula:
\begin{eqnarray}
\label{TrES-2}
\D v=-fv_p\left[1.06-0.0754\left(\frac{v_p}{v\sin i_\star}\right)^2\right].
\end{eqnarray}
Again, in Figure \ref{tres4-fig} (b), crosses indicate
the simulated result (eq.[\ref{TrES-2}]).
We also show the OTS formula and the Gaussian formula (\ref{gaussianDv}),
in which $\sigma=v\sin i_\star/1.30=0.766$ km sec$^{-1}$.
We here adopt $i_o=83.6^\circ$ and $\la=0^\circ$.

The three different lines are almost
indistinguishable due to the small RM amplitude, but we can safely say that
the OTS formula as well as the Gaussian formula (\ref{gaussianDv}) well
	   describe
the simulated result. When the stellar spin velocity is small enough, the OTS
formula provides a good approximation to the velocity anomaly.
\item[TrES-4] TrES-4 is a transiting planetary system discovered by
\citet{Mandushev2007}.  The parent star is a F-type star with an
effective temperature of $6200\pm75$ K and has a relatively high rotation
velocity of $v\sin i_\star=8.5\pm0.5$ km~sec$^{-1}$
\citep{Sozzetti2009}.  Since the spectral type of the star is different
from that of the Sun, we use theoretically synthesized F-type spectrum.
We then broaden the theoretical spectrum assuming a quadratic
limb-darkening law (eq. [\ref{limbdark}]) with $u_1=0.46$ and $u_2=0.31$
\citep{Claret2004}, and obtain
\begin{eqnarray}
\label{TrES-4}
\D v=-fv_p\left[1.62-0.885\left(\frac{v_p}{v\sin
			    i_\star}\right)^2\right].
\end{eqnarray}
Figure \ref{tres4-fig} (c) shows the result.  The
misalignment angle $\la$ and the orbital inclination $i_o$ are assumed
to be $0^\circ$ and $82.6^\circ$, respectively.  
The solid line in Figure \ref{tres4-fig} indicates the Gaussian formula with 
$\sigma = v\sin i_\star/1.32\approx
6.4$ km sec$^{-1}$.

\item[HD209458] 
Finally, we compare the Gaussian formula with the simulated 
radial velocity anomaly of HD209458 system.
While in Figure \ref{nf2} the Gaussian broadening kernel is applied to synthesize
 the mock spectra for stellar rotation, we 
here adopt the actual rotational kernel (eq.[\ref{rotation}]) with $u_1=0.45$, 
$u_2=0.30$, and $v\sin i_\star=4.5$ km sec$^{-1}$. 
Here is the empirical formula after the mock analysis:
\begin{eqnarray}
\label{HD209458}
\D v=-fv_p\left[1.49-0.684\left(\frac{v_p}{v\sin i_\star}\right)\right].
\end{eqnarray}
As in the above three cases, we show the OTS formula and the Gaussian 
formula  (\ref{gaussianDv}) in Figure \ref{tres4-fig} (d). We adopt $\sigma=v\sin i_\star/1.31
=3.4$ km sec$^{-1}$, $i_o=86.7^\circ$, and $\la=0^\circ$.

\end{description}

The Gaussian formulae in Figure \ref{tres4-fig} well describe the
behavior of simulated results in each system even though the actual
rotation kernels are not described by simple Gaussians. 
Comparing the three systems except TrES-2 system, 
in which the stellar velocity is particularly small, we notice that the Gaussian 
formulae for TrES-4 and HD209458 describes the simulation better than that for HD17156.
This is presumably due
to the fact that apart from the tiny contribution of the tail, the
rotation kernel is better approximated as a Gaussian for the star with
higher rotational velocity (Appendix \ref{s:leastsquare}).  The fact
that the three lines in each figure intersect at the same time (close
to, but not identical to the central transit time time $=0$) indicates
that while the difference among the OTS formula, the Gaussian formula
and the simulated result affects the estimate of $v\sin i_\star$, the
estimated value of the spin-orbit angle is fairly robust as discussed
in \S2.

We also notice a fairly big discrepancy between the Gaussian formula
and the simulated result during the egress of HD17156 (at $0<$ time[h]
$<1.5$) in the upper panel of Figure \ref{tres4-fig}. Indeed the OTS
formula approximates better the simulated result at the egress phase.
In order to understand the behavior, we systematically changed the
value of $i_o$, and found that the asymmetry of the velocity anomaly
curve between ingress and egress is very sensitive to the combination
of $\lambda$ and $i_o$. This points to a limitation of our Gaussian
approximation, which might be fixed by using a more precise rotation
kernel.


\section{Summary and Discussion \label{s:sum}}

We have presented an improved analytic formula for the RM effect as 
measured from the cross-correlation method. Our main finding is equation
(\ref{gaussianDv}) which describes the RM velocity anomaly under the
Gaussian approximation for the intrinsic line profile and the stellar
rotation kernel. Unlike the previous approximation (the moment method)
adopted by OTS, the radial velocity anomaly from the cross-correlation
method is explicitly dependent both on the thermal and natural
broadening width of an individual line profile ($\beta$) and on the
rotation velocity ($\sigma$).

The analytical formula has been compared with the simulated analysis
for the Subaru HDS routine using the mock spectra constructed from the
high-resolution spectrum of the Sun. Even though the actual analysis
routine attempts to fit simultaneously many different lines that are not
necessarily approximated by Gaussians, the resulting RM velocity anomaly
is well described by a form of equation (\ref{gaussianDv}).

The current result explains the previous findings
\citep{Winn2005,Johnson2008,Winn2008} that the OTS formula does not
provide a good approximation to planetary systems with a relatively
high stellar rotation rate. The next step in obtaining a more accurate
analytic formula for the RM effect would probably be to use a more
accurate rotation kernel (eq.~[\ref{rotation}]) instead of the
Gaussian approximation (\ref{eq:RapproxG}). For future work we plan to
pursue this approach, together with the propagation of the resulting
precision on the misalignment angle $\lambda$ and the rotation
velocity $v\sin i_\star$.

In addition to the improvement of the analytic approach, the current
result points to a possible further refinement of the observational
reduction routine for the velocity anomaly due to the RM effect.
The procedure described in \S 3.1 is appropriate for the radial velocity
determination {\it out of transit}. During the transit, however, 
the Doppler-shifted stellar spectrum $S(\la-\d\la)$ in equation
(\ref{model}) should be replaced with
\begin{eqnarray}
&S_{\mathrm{star}}(\la-\tilde{\la})-S_{\mathrm{planet}}(\la-\tilde{\la}-\D\la)
\nonumber\\
=&S_{\mathrm{star}}(\la-\tilde{\la})
-fS_{\mathrm{narrowed~ template}}(\la-\tilde{\la}-\D\la),
\end{eqnarray}
where $S_{\mathrm{star}}(\la)$ is the stellar template spectrum derived
by the procedure described in \S3.1 during the out-of-transit data,
$S_{\mathrm{narrowed~ template}}(\la)$ is obtained by deconvolving
$S_{\mathrm{star}}(\la)$ with the rotational kernel in principle, and
$\tilde{\la}$ and $\D\la$ correspond to the Keplerian plus peculiar
velocity of the star and the radial velocity of the occulted portion of
the stellar disk, respectively.

Fitting the observed spectrum {\it in transit} with the additional
parameters $f$, $\tilde{\la}$, and $\D\la$, instead of $\d\la$ alone, one
may directly extract the velocity $v_p= c\D\la/\la_0$ after averaged
over many different lines.  Moreover one may reduce the number of
fitting parameters by estimating $f$ from photometric data during the
transit. 

Since the above method is more appropriate to model the spectra of the
transiting planetary system during transit, one expects that the
precision and accuracy of $\D\la$, and therefore the RM velocity
anomaly, may be improved in principle.  In reality, however, the
practical feasibility to implement in the analysis routine depends on the
available resolution of the spectra as well as their stability. The work
toward this improvement is also currently under way, which we also hope
to present elsewhere in due course.

The RM effect has become an observationally mature probe of the
transiting planetary systems. Further improvement in both the
accuracy and the precision of the velocity anomaly is important to
advance the understanding of the formation and evolution processes of
such systems. We hope that the present analytic approach provides some
insights that will enable progress in that direction.

\acknowledgments 

We thank Wako Aoki for useful advice on the spectral line profiles.
The radial velocity analysis in our simulation was carried out on the
computer system at the Astronomy Data Center of the National
Astronomical Observatory of Japan.  N.N. is supported by a Japan
Society for Promotion of Science (JSPS) Fellowship for Research (PD:
20-8141).  Y.S. gratefully acknowledges the support from the Global
Scholars Program of Princeton University.  This work is also supported
by JSPS Core-to-Core Program ``International Research Network for Dark
Energy''. J.N.W.\ gratefully acknowledges the support of the NASA
Origins program through award NNX09AD36G.

\appendix

\section{Flux during Complete Transit \label{s:flux}}

In this Appendix, we compute the flux ratio between the flux coming from the
stellar disk blocked by the planet and the total flux, and show that 
equation (\ref{Dv}) reproduces the OTS formula.

In the case of quadratic limb darkening (eq. [\ref{limbdark}]), 
we can express $f$ in terms of $\gamma$, $u_1$, $u_2$, and the position 
of the planet in the stellar disk. The definition of $f$ is
\begin{eqnarray}
f(x, y)\equiv \frac{\displaystyle\int_{\mathrm{planet}} I(x^\prime, y^\prime)
dx^\prime dy^\prime}{\displaystyle\int_{\mathrm{stellar~disk}} 
I(x^\prime, y^\prime)dx^\prime dy^\prime},\label{fluxdef}
\end{eqnarray}
where  the $x$- and $y$-axes are defined so that the $y$-axis is along 
the projected rotational axis and the origin is at the center of the stellar disk. 
The integration in the numerator covers the portion blocked by planet, 
while the one in the denominator is performed throughout the whole stellar disk.
Denoting the radius of the star by $R_s$, then the cosine in equation 
(\ref{limbdark}) is written as
\begin{eqnarray}
\cos\theta=\sqrt{1-\frac{x^2+y^2}{R_s^2}}\equiv\sqrt{1-\frac{r^2}{R_s^2}}.
\end{eqnarray}
Then the denominator of equation (\ref{fluxdef}) becomes
\begin{eqnarray}
&&\int_0^{R_S}dr2\pi r\left[1-u_1\left\{1-\sqrt{1-\left(\frac{r}{R_s}\right)^2}\right\}
-u_2\left\{1-\sqrt{1-\left(\frac{r}{R_s}\right)^2}\right\}^2\right]\nonumber\\
&=&\pi R_s^2(1-\frac{1}{3}u_1-\frac{1}{6}u_2).\label{deno1}
\end{eqnarray}
When the size of the planet is sufficiently small, the numerator of 
equation (\ref{fluxdef}) can be approximated as
\begin{eqnarray}
\int_{\mathrm{planet}} I(x^\prime, y^\prime)dx^\prime dy^\prime&\simeq
& I(x, y)\pi R_s^2\gamma^2\nonumber\\
&=&\pi R_s^2\gamma^2\left[1-u_1\left\{1-\sqrt{1-\rho^2}\right\}-u_2
\left\{1-\sqrt{1-\rho^2}\right\}^2\right],\label{nume1}
\end{eqnarray}
where $\rho\equiv\sqrt{x^2+y^2}/R_s$ and $\gamma=R_p/R_s$. 
Substituting equations (\ref{deno1}) and (\ref{nume1}) into equation (\ref{fluxdef}), 
we obtain
\begin{eqnarray}
f(x, y)\simeq\frac{1-u_1\left\{1-\sqrt{1-\rho^2}\right\}-u_2\left\{1-\sqrt{1-\rho^2}\right\}^2}
{1-u_1/3-u_2/6}\gamma^2.\label{f}
\end{eqnarray}
With this expression, equation (\ref{Dv}) reduces to
\begin{eqnarray}
\D v\simeq -v_p\frac{\gamma^2\left[1-u_1\left\{1-\sqrt{1-\rho^2}\right\}
-u_2\left\{1-\sqrt{1-\rho^2}\right\}^2\right]}{(1-u_1/3-u_2/6)
-\gamma^2\left[1-u_1\left\{1-\sqrt{1-\rho^2}\right\}-u_2\left\{1-\sqrt{1-\rho^2}\right\}^2\right]},\label{limbOTS}
\end{eqnarray}
where $v_p$ is expressed by the position of the planet as
\begin{eqnarray}
v_p (x, y)=\frac{\displaystyle{\iint_{\mathrm{planet}}\Omega_s(x^\prime, y^\prime)
x^\prime\sin i_\star~dx^\prime dy^\prime}}{\displaystyle\iint_{\mathrm{planet}}dx^\prime dy^\prime}.\label{subp}
\end{eqnarray}
The function 
$\Omega_s(x, y)$ is the angular velocity of the stellar spin at position $(x, y)$ 
and $i$ is the inclination of the stellar spin axis. 
For rigidly rotating stars, the planet velocity during a complete transit becomes
\begin{eqnarray}
v_p=\Omega_sx\sin i_\star,\label{subp3}
\end{eqnarray}
with $\Omega_s$ being a constant.
Equation (\ref{limbOTS}) with (\ref{subp3}) reproduces the OTS formula for 
complete transit with limb-darkening (eq. [40] in OTS). 

\section{During Ingress and Egress \label{s:ingress}}

During an ingress and an egress of transits, the fraction of the portion
on the stellar disk occulted by the planet changes.  The center of the
stellar disk and the planet is at the origin and at $(x, y)$,
respectively.
\begin{figure}[h]
\begin{center}
\includegraphics[width=10cm]{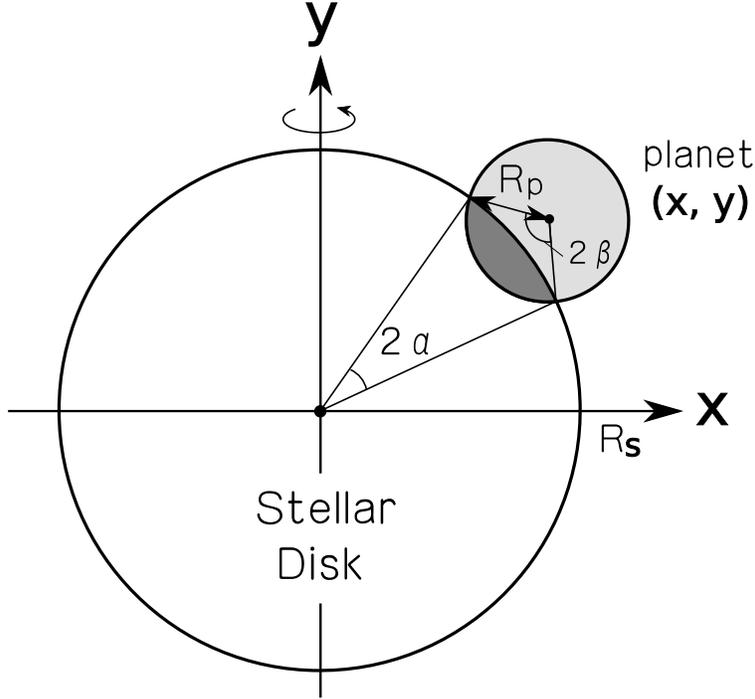}
\caption{Schematic illustration of the configuration at an ingress or
egress. The center of the planet is at $(x, y)$. The dark shaded region
is occulted by the planet.  }\label{ingress}
\end{center}
\end{figure}

Figure \ref{ingress} shows an example of a configuration during an
ingress or an egress. We define the radius of the planet as $R_p$. The
angles $\alpha$ and $\beta$ are given by
\begin{eqnarray}
\cos\alpha=\frac{R_s^2+r^2-R_p^2}{2R_sr},~
\cos\beta=\frac{R_p^2+r^2-R_s^2}{2R_pr},
\end{eqnarray}
respectively.
The size $S$ of the dark shaded area in Figure \ref{ingress} (the portion 
occulted by the planet) is 
\begin{eqnarray}
S=\alpha R_s^2-\frac{1}{2}R_s^2\sin2\alpha+\beta R_p^2-\frac{1}{2}R_p^2\sin2\beta.
\end{eqnarray}
Thus, the fraction $\Gamma(r)$ of the portion on the stellar photosphere 
occulted by the planet becomes
\begin{eqnarray}
\Gamma(r)\equiv \frac{S}{\pi R_s^2}=\frac{1}{\pi}\left\{\alpha-\frac{1}{2}
\sin2\alpha+\gamma^2(\beta-\frac{1}{2}\sin2\beta)\right\}.
\end{eqnarray}
During an ingress or an egress, $\gamma^2$ in equation (\ref{f}) must be 
replaced with $\Gamma(r)$ depending on the position of the planet. 
This result is derived in the case of $R_s<r$, but we can prove this also 
holds in the case of $R_s\geq r$.

The subplanet velocity is no longer expressed by equation (\ref{subp3}). 
In this case, we must replace equation (\ref{subp3}) with 
\begin{eqnarray}
v_p (x, y)=\frac{\displaystyle{\iint_{S}\Omega_sx^\prime\sin i_\star~dx^\prime dy^\prime}}
{\displaystyle\iint_{S}dx^\prime dy^\prime},\label{subp2}
\end{eqnarray}
where $S$ is the dark shaded region in Figure \ref{ingress}. 
However, as long as the planet is sufficiently small enough, 
the approximation (\ref{subp3}) provides a good description for 
subplanet velocity (\ref{subp2}).

\section{Derivation of Equation (\ref{exact})}
\label{s:derivation}
We here present the derivation of equation (\ref{exact}) in detail.
Below, we implicitly assume that the quantities $\beta_\star$ and 
$\gamma_\star$ represent the broad line originated from the whole 
photosphere, while $\beta_p$ and $\gamma_p$ indicate the 
narrow line originated from the portion of the planet. 
For further computation, it is convenient to consider the Fourier 
transforms of the line profile functions.
Note that the Fourier transforms of Gaussian and Lorentzian 
respectively become
\begin{eqnarray}
G(\la)=\frac{1}{\beta\sqrt{\pi}}e^{-\la^2/\beta^2}&\Longrightarrow& 
\tilde{G}(\s)=e^{-\pi^2\beta^2\s^2},\label{gaussfourier}\\
L(\la)=\frac{1}{\pi}\frac{\gamma^2}{\la^2+\gamma^2}&\Longrightarrow& 
\tilde{L}(\s)=e^{-2\pi \gamma |\s|}.\label{lorentzfourier}
\end{eqnarray}
Then, the Fourier transform of the Voigt function becomes $\tilde{V}(\s)
=\tilde{G}(\s)\ti{L}(\s)$. In terms of these, equations (\ref{F1}) and (\ref{F2}) 
are rewritten as
\begin{eqnarray}
&\F_\mathrm{transit}(\la)=-\displaystyle{\integ} d\s \left[\ti{V}_\star(\s)e^{-2\pi i\s \la}
-f\ti{V}_p(\s)e^{-2\pi i\s(\la-\D\la)}\right],\\
&\F_\mathrm{star}(\la)=-\displaystyle\integ d\s\ti{V}_\star(\s)e^{-2\pi i\s(\la)},
\end{eqnarray}
where $\ti{V}_\star(\sigma)$ and $\ti{V}_p(\sigma)$ are the Fourier transforms 
of $G(\la;\beta_\star)L(\la;\gamma_\star)$ and $G(\la;\beta_p)L(\la;\gamma_p)$, 
respectively.
The corresponding Fourier counterpart are 
\begin{eqnarray}
&\ti{\F}_\mathrm{transit}(\s)=-\ti{V}_\star(\s)+f \ti{V}_p(\s)e^{2\pi i\s \D\la}\\
&\ti{\F}_\mathrm{star}(\s)=-\ti{V}_\star(\s).
\end{eqnarray}
Thus, we obtain
\begin{eqnarray}
C(\d_\mathrm{RM})=\integ d\s \ti{C}(\s) e^{2\pi i\s\d_\mathrm{RM}},
\end{eqnarray}
where
\begin{eqnarray}
\ti{C}(\s)&=&\ti{\F}_\mathrm{star}(\s)\times \ti{\F}^*_\mathrm{transit}(\s)\nonumber\\
&=&\tilde{V}_\star(\s)\left(\ti{V}_\star(\s)-f\ti{V}_p(\s)e^{-2\pi i \s \D\la}\right).\label{tildeC}
\end{eqnarray}
Now it is easy to incorporate additional effects expressed as a convolution 
simply by multiplying the Fourier transform of the additional convolution kernel.

Substituting equations (\ref{gaussfourier}) and (\ref{lorentzfourier}) into 
equation (\ref{tildeC}), we obtain
\begin{eqnarray}
C(\s)&=& \integ d\s \exp \left[-2\pi^2\beta_\star^2\s^2-4\pi\gamma_\star|\s|
+2\pi i \s\d_\mathrm{RM}\right]\nonumber\\
&&-f\integ d\s \exp \left[-\pi^2(\beta_\star^2+\beta_p^2)\s^2
-2\pi(\gamma_\star+\gamma_p)|\s|+2\pi i \s(\d_\mathrm{RM}-\D\la)\right].\label{inverse}
\end{eqnarray}
The integral of each term in equation (\ref{inverse}) results in the 
following integral:
\begin{eqnarray}
\xi(a,b,c)\equiv\int_0^{\infty} \exp (-bx^2+cx)\cos(ax)dx.\label{xidef}
\end{eqnarray}
The derivative of the function $\xi(a,b,c)$ with respect to $a$ is 
\begin{eqnarray}
\frac{\partial}{\partial a}\xi(a,b,c)=-\frac{1}{8b^{3/2}}e^{-(a^2-c^2)/4b}
\sqrt{\pi}\left[(a-ic)e^{iac/2b}\left(1+\mathrm{erfi}(\frac{a-ic}{2\sqrt{b}})\right)
+\mathrm{c.c.}\right].\label{partial}
\end{eqnarray}
In terms of $\xi(a,b,c)$, equation (\ref{inverse}) is expressed as
\begin{eqnarray}
C(\s) = \xi\left(2\pi \d_\mathrm{RM}, 2\pi^2\beta_\star^2, 4\pi\gamma_\star\right) 
-f\xi \left(2\pi (\d_\mathrm{RM}-\D\la), \pi^2(\beta_\star^2+\beta_p^2), 
2\pi(\gamma_\star+\gamma_p)\right). \label{xi}
\end{eqnarray}
Substituting equation (\ref{xi}) into equation (\ref{deriva}) and using 
equation (\ref{partial}), we finally obtain an equation for $\d_\mathrm{RM}$:
\begin{eqnarray}
&\displaystyle\frac{1}{b^{3/2}}e^{-(a^2-c^2)/4b}\mathrm{Re}
\left[(a-ic)e^{iac/2b}\left(1+\mathrm{erfi}(\frac{a-ic}{2\sqrt{b}})\right)\right]\nonumber\\
&=\displaystyle \frac{f}{B^{3/2}}e^{-(A^2-C^2)/4B}\mathrm{Re}
\left[(A-iC)e^{iAC/2B}\left(1+\mathrm{erfi}(\frac{A-iC}{2\sqrt{B}})\right)\right],\label{exact2}
\end{eqnarray}
where 
\begin{eqnarray}
&a=2\pi \d_\mathrm{RM}, ~
b=2\pi^2\beta_\star^2,~ 
c=4\pi\gamma_\star,\\
&A=2\pi(\d_\mathrm{RM}-\D\la),~
B=\pi^2(\beta_\star^2+\beta_p^2),~
C=2\pi(\gamma_\star+\gamma_p).
\end{eqnarray}

\section{Lorentzian profile}
\label{s:lorentziansection}
In \S2.3, we have derived analytic expression for Gaussian line profiles.
In this appendix, we consider the opposite simplification that line profile 
has a Lorentzian form ($b=B=0$). To derive the radial velocity formula, we use
\begin{eqnarray}
\lim_{b\to0}\displaystyle\frac{1}{b^{3/2}}e^{-(a^2-c^2)/4b}\mathrm{Re}
\left[(a-ic)e^{iac/2b}\left(1+\mathrm{erfi}(\frac{a-ic}{2\sqrt{b}})\right)\right]
=-\frac{8}{\sqrt{\pi}}\frac{ac}{(a^2+c^2)^2},
\end{eqnarray}
Then, equation (\ref{exact}) or (\ref{exact2}) reduces to 
\begin{eqnarray}
\frac{2\d_\mathrm{RM}\gamma_\star}{(\d_\mathrm{RM}^2+4\gamma_\star^2 )}
=f\frac{(\d_\mathrm{RM}-\D\la)(\gamma_\star+\gamma_p)}{\{ (\d_\mathrm{RM}-\D\la)^2
+(\gamma_\star+\gamma_p)^2\}^2}.
\end{eqnarray}
Employing the similar approximation in \S 2.3.1, we obtain an approximate 
expression for Lorentzian profile: 
\begin{eqnarray}
\d_\mathrm{RM}\simeq -\left(\frac{2\gamma_\star}{\gamma_\star+\gamma_p}\right)^3 
f \D \la\left\{1-2\left(\frac{\D\la}{\gamma_\star+\gamma_p}\right)^2\right\},
\end{eqnarray}
or
\begin{eqnarray}
\D v\simeq-\left(\frac{2\gamma_\star}{\gamma_\star+\gamma_p}\right)^3 
f v_p\left\{1-\frac{2\la_0^2}{c^2(\gamma_\star+\gamma_p)^2}v_p^2\right\}.\label{lorentzianDv}
\end{eqnarray}
Again, we obtain the similar expression to equation (\ref{gaussianDv}). 
Note that the factor in front of $f$ depends on the widths of Lorentzians, 
which generally differs from unity. The presence of the cubic term in $v_p^3$ 
is also a major difference from the result by moment-method.

\section{Velocity Anomaly for the Gaussian Profile Perturbed with Lorentzian}
\label{s:perturbation}
We discuss a more general case in which the line profile is expressed as
Gaussian profile with a small contribution of Lorentzian profile.
We assume that $\gamma/\beta$ is sufficiently small 
(the Gaussian property is dominant) and
 collect the terms up to the linear order of $\gamma/\beta$ in 
 equation (\ref{exact}). Noting that 
\begin{eqnarray}
\mathrm{erfi}(z)=\frac{2z}{\sqrt{\pi}} ~~~~~~~z\ll 1,
\end{eqnarray}
and defining $z=(a-ic)/2\sqrt{b}$, the right hand side of equation 
(\ref{exact}) becomes
\begin{eqnarray}
\frac{2}{b}\mathrm{Re}\left[ze^{-z^2}\left(1+\frac{2i}{\sqrt{\pi}}z\right)\right]&\simeq
& \frac{2}{b}\mathrm{Re}\left[z(1-z^2)\left(1+\frac{2i}{\sqrt{\pi}}z\right)\right]\nonumber\\
&\simeq & \frac{2a}{b\sqrt{b}}\left(\frac{1}{2}+\frac{c}{\sqrt{\pi b}}-\frac{a^2-3c^2}{8b}\right).
\end{eqnarray}
Expanding the left hand side of equation (\ref{exact}) in a similar 
way, and 
substituting $a,b,c$ and $A,B,C$, we obtain
\begin{eqnarray}
\d_\mathrm{RM} \simeq -\left(\frac{2\beta_\star^2}{\beta_\star^2+\beta_p^2}\right)^{3/2}
\frac{f\D\la}{1+\frac{8\gamma_\star}{\sqrt{2\pi\beta_\star^2}}}
\left\{1+\frac{4(\gamma_\star+\gamma_p)}{\sqrt{\pi(\beta_\star^2+\beta_p^2)}}
-\frac{\D\la^2}{\beta_\star^2+\beta_p^2}\right\},
\end{eqnarray} 
or
\begin{eqnarray}
\D v\simeq -\left(\frac{2\beta_\star^2}{\beta_\star^2+\beta_p^2}\right)^{3/2}
\frac{1}{1+\frac{8\gamma_\star}{\sqrt{2\pi\beta_\star^2}}}fv_p
\left\{1+\frac{4(\gamma_\star+\gamma_p)}{\sqrt{\pi(\beta_\star^2+\beta_p^2)}}
-\frac{\la^2}{c^2(\beta_\star^2+\beta_p^2)}v_p^2\right\}.\label{analyticvoigt}
\end{eqnarray}
This is the formula for the velocity anomaly when absorption lines are 
Gaussians perturbed with Lorentzian profile. Note that equation 
(\ref{analyticvoigt}) reproduces equation (\ref{gaussianDv}) 
when we set $\gamma_\star=\gamma_p=0$.

\section{Relation between the Rotation and Gaussian 
Broadening Kernel by Least-squares Fitting}
\label{s:leastsquare}
Here, we derive a scaling factor between the upper-limit 
wavelength $\la_L$ of the rotational kernel and the dispersion 
$\sigma$ of the Gaussian broadening kernel by least-squares fitting.
Although replacing a rotational kernel with Gaussian one 
is a crude approximation, 
we can justify this treatment to some extent (\S 3.3.2.).

We define $\D^2$ as the residual between the two convolution kernels:
\begin{eqnarray}
\D^2&=&\integ [R(\la;\la_L)-G(\la;\beta)]^2d\la\nonumber\\
&=&\integ R^2(\la;\la_L)d\la + \integ G^2(\la;\beta)d\la -2 \integ 
R(\la;\la_L)G(\la;\beta)d\la.\label{delta^2}
\end{eqnarray}
The scaling factor $\alpha$ is define as 
\begin{equation}
\alpha\equiv\frac{\la_L}{\beta}.
\end{equation}
When we fix the rotational velocity (thus $\la_L$), $\alpha$ is determined 
by minimizing of the function $\D^2$:
\begin{equation}
\frac{\partial \D^2}{\partial \alpha}=0.
\end{equation}
Since the first term in equation (\ref{delta^2}) does not depend on $\beta$,  
we neglect it in computing the derivative in terms of $\beta$. 
The second term is written as
\begin{eqnarray}
\integ G^2(\la;\beta)d\la = \frac{1}{\beta\sqrt{2\pi}}=\frac{\alpha}{\la_L\sqrt{2\pi}}.
\end{eqnarray}
The third term in equation (\ref{delta^2}) reduces to
\begin{eqnarray}
\label{eq:F1-3}
\integ R(\la;\la_L)G(\la;\beta)d\la&=&\frac{\la_L}{\beta\sqrt{\pi}}
\left[c_1\int_{-1}^1e^{-\a^2x^2}\sqrt{1-x^2}dx\right.\nonumber\\
&&\left. +c_2\int_{-1}^1e^{-\a^2x^2}(1-x^2)dx
+c_3\int_{-1}^1e^{-\a^2x^2}(1-x^2)^{3/2}dx\right]\nonumber\\
&=&\frac{\a}{\sqrt{\pi}}\left[c_1\frac{\pi}{2}e^{-\a^2/2}\left\{I_0(\a^2/2)+I_1(\a^2/2)\right\}\right.\nonumber\\
&&+c_2\frac{1}{2\a^3}\left\{2\a e^{-\a^2}+(-1+2\a^2)\sqrt{\pi}~\mathrm{erf}(a)\right\}\nonumber\\
&&\left.+c_3\frac{\pi}{2\a^2}e^{-\a^2/2}\left\{\a^2I_0(\a^2/2)+(-1+\a^2)I_1(\a^2/2)\right\}
\right],
\end{eqnarray}
where $I_n(x)$ is the $n$-th order modified Bessel function. 
Substituting equations (\ref{c_1}), (\ref{c_2}) and (\ref{c_3}) into
equation (\ref{eq:F1-3}), we  obtain the 
derivative of $\D^2$:
\begin{eqnarray}
\displaystyle\frac{\partial \D^2}{\partial \a}&=&\frac{1}{\la_L\sqrt{\pi}}
\left[\frac{1}{\sqrt{2}}-\frac{12(u_1+2u_2)e^{-\a^2}}{\a^2(-6+2u_1+u_2)}\right.\nonumber\\
&&+\frac{6e^{-\a^2/2}}{\a^3(-6+2u_1+u_2)}\left\{-2\a^3(-1+u_1+u_2)I_0(\a^2/2)\right.\nonumber\\
&&+2\a\{-2u_2+\a^2(-1+u_1+u_2)\}I_1(\a^2/2)\nonumber\\
&&\left.\left.+\sqrt{\pi}e^{\a^2/2}(u_1+2u_2)\mathrm{erf}(\a)\right\}\right]=0.\label{Delta^2}
\end{eqnarray}
Substituting the limb-darkening parameters, we numerically solve the
above equation for $\a$. We obtain $\a\simeq1.31$ for HD17156
($u_1=0.6$, $u_2=0$), $\a\simeq1.30$ for TrES-2 ($u_1=0.40$,
$u_2=0.30$), $\a\simeq1.32$ for TrES-4 ($u_1=0.46$,
$u_2=0.31$), and $\a\simeq1.31$ for HD209458 ($u_1=0.45$,
$u_2=0.30$).  
Equation (\ref{Delta^2}) implies that residuals $\D^2$ scales as 
$\la_L^{-1}$ when we fix $\alpha$.
This means that for rapidly rotating star, $\D^2$ becomes small and 
the approximation of rotational kernel with Gaussian is validated.



\end{document}